\documentclass{moriond}

\usepackage{amsmath}

\def\apj{{\em ApJ}}                 
\def\cqg{{\em Class. Quant. Grav.}} 
\def\prd{{\em Phys.~Rev.~D}}        
\def\prl{{\em Phys.~Rev.~Lett.}}    
\def\mnras{{\em MNRAS}}             

\def\be{\begin{equation}}
\def\ee{\end{equation}}
\def\bea{\begin{eqnarray}}
\def\eea{\end{eqnarray}}

\newcommand{\Photo}{}

\begin{document}
\vspace*{4cm}
\title{STATUS OF THE CONTINUOUS GRAVITATIONAL WAVE SEARCHES IN THE ADVANCED DETECTOR ERA}

\author{ M. Bejger (for the LIGO-Virgo Collaboration)} 

\address{Nicolaus Copernicus Astronomical Center, Polish Academy of Sciences,
ul. Bartycka 18, 00-716, Warszawa, Poland} 

\maketitle
\abstracts{Periodic (almost monochromatic) gravitational waves emitted by
rotating, asymmetric neutron stars are intriguing potential signals in the
sensitivity band of Advanced LIGO and Advanced Virgo detectors. These signals
are related to elastic and magnetic stresses in the neutron-star interior, as
well as to various possible instabilities, and thus are interesting from the
point of view of the largely-unknown neutron star structure. I will describe
the main challenges related to these searches, the current state of the
data-analysis methods and plans for the future.} 

\section{Introduction}

Recent first direct detections of gravitational waves with the two LIGO
detectors \cite{GW150914,GW151226} create an unprecedented opportunity for
studying the Universe through a novel, never before explored channel of
spacetime fluctuations. Advanced LIGO \cite{ALIGO2015} and Advanced Virgo
\cite{AdV2015}, few-kilometer long arm laser interferometric detectors are
sensitive in the range of frequencies between 10 Hz and a few kHz.  They
registers a coherent signal emitted by a bulk movement of large, rapidly-moving
masses. Once emitted, gravitational waves are weakly coupled to the surrounding
matter and propagate freely without scattering. This has to be contrasted with
the electromagnetic emission which originates at the microscopic level, is
strongly coupled to the surroundings and often reprocessed; it carries a
reliable information from the last scattering surface only. Gravitational wave
observations are therefore the perfect counterpart to the electromagnetic
observations as they may provide us with information impossible to obtain by
other means.

In addition to inspiralling and merging binary systems, among promising 
sources of gravitational radiation are all asymmetric collapses
and explosions e.g., supernov{\ae}. wide binary systems, rotating deformed
stars (gravitational-wave `pulsars' of continuous and transient nature), as
well as stochastic background waves produced by whole populations of sources. 

In the following we will briefly describe astrophysical motivation behind
continuous gravitational waves produced by rotating deformed neutron stars
(Sect.~\ref{sect:motiv}), data-analysis methods and computational challenges
related to them (Sect.~\ref{sect:meth}), Advanced LIGO O1 run results
(Sect.~\ref{sect:curr}), and plans for the future (Sect.~\ref{sect:plans}). 

\section{Astrophysical motivation} \label{sect:motiv}

Neutron stars are the most relativistic, dense and compact {\em material}
objects in the Universe.  Their compactness i.e., mass-to-radius ratio
$2GM/Rc^2$ reaches 0.5 ($G$ being the Newton's constant, $c$ the speed of
light); for comparison, the compactness of the most compact objects, black
holes, equals 1. Their average density surpasses the nuclear saturation density
i.e., the density of atomic nuclei.  At these densities matter exists most
probably in an `exotic' phase, e.g., as de-confined quarks. Neutron stars are
self-gravitating objects stabilized by strong force interactions, gigantic
nuclei of masses up to $2\,M_\odot$, radii about $10-15$ km with surface
magnetic fields of $10^8-10^{15}$ Gauss, that can spin several hundred times
per second. Comparison of realistic models of neutron stars with a variety of
astrophysical observations - including the gravitational waves they emit - is
the only way to peek into a realm of dense matter strong interactions much
above the nuclear saturation density. 

Neutron stars provide truly unique conditions to study matter at the most
extreme densities, pressures, and in the presence of powerful magnetic fields.
These conditions cannot be reproduced (or even approximated) in terrestrial
laboratories. At present, about 2500 neutron stars are known, and an estimated
number of $10^8-10^9$ exists in every galaxy similar to ours. Neutron stars
play an important role in many astrophysical phenomena: they are observed in
all the EM spectrum as radio-, X- and $\gamma$-pulsars, magnetars, are present
in supernov{\ae} remnants, in many accreting systems and in relativistic double
neutron star binaries, yet very little is known about their internal
composition. What is conventionally accepted is that at least some part of
neutron-star interior - the outer part about 1 km thick called the crust,
corresponding to densities {\em below} the nuclear saturation density - is in
the crystalline state. 

In the following we will focus on rotating, non-axisymmetric neutron stars as
sources of continuous periodic gravitational wave emission. Continuous
gravitational wave is by definition a long-lived phenomenon, $T > T_{obs}$, and
its frequency $f_{GW}$ is somehow proportional to the spin frequency of the star $f$,
$f_{GW}\propto f$. There are several proposed astrophysical mechanisms
providing the necessary asymmetry, which in the case of a rotating star is the
source of time-varying quadrupole required for the gravitational-wave emission.
Mechanisms include neutron-star ''mountains'', supported by elastic and/or
magnetic stresses ($f_{GW} = 2f$), oscillations (e.g., r-modes
\cite{AnderssonK2001}, $f_{GW} = 4/3f$), free precession ($f_{GW}\propto
f + f_{prec}$) and accretion that drives the deformation from r-modes,
thermal gradients and magnetic fields ($f_{GW}\simeq f$). For a recent
review see \cite{Lasky2015}. 

The most-commonly used and the simplest model of the non-axisymmetric rotating
neutron star radiating purely quadrupolar waves consists of a triaxial
ellipsoid (with moments of inertia along the axes $I_1$, $I_2$, $I_3$),
rotating about one of the principal directions of its moment of inertia tensor
($I_3$, say). Such a body radiates GWs at the frequency twice the rotational
frequency of the star, $2\pi f_{GW} = \Omega_{GW} = 2\Omega$. The strain signal
at the detector changes in time as  
\begin{equation} 
h(t) = h_0\left( \frac{1}{2}F_+(t,\alpha,\delta,\psi)\left(1+\cos^2\iota\right)\cos(\phi(t) + \phi_0) 
+ F_\times(t,\alpha,\delta,\psi)\cos\iota\cos(\phi(t) + \phi_0)\right), 
\label{eq:triaxial} 
\end{equation} 
where $h_0$ is the gravitational-wave strain amplitude, $\alpha$ and $\delta$
are right ascension and declination of the source in the sky, $\psi$ is the
polarization angle, and $\iota$ the inclination of the rotation axis to the
line of sight. Phase of the signal $\phi(t) + \phi_0$ incorporates the possible
evolution of the spin frequency. $F_+$ and $F_\times$ are the antenna responses
of the detector, corresponding to two gravitational-wave polarizations $+$ and
$\times$.  

From the quadrupole formula \cite{Einstein1918}, amplitude $h_0$ is estimated as follows:
\begin{eqnarray}
h_0 = \frac{16\pi^2G}{c^4}\frac{I\epsilon f^2}{d} = 
4\times 10^{-25}\left(\frac{\epsilon}{10^{-6}}\right)
\left(\frac{I}{10^{45}\ {\rm g\ cm^2}}\right)
\left(\frac{f}{100\ {\rm Hz}}\right)^2
\left(\frac{100\ {\rm pc}}{d}\right), 
\label{eq:h0}
\end{eqnarray}
where $I\equiv I_3$, $f=\Omega/2\pi$, $\epsilon = (I_1 - I_2)/I$ is is the
fiducial equatorial ellipticity of the star (a ''deformation''), and $d$ is
typical distance in the Galaxy.  According to theoretical studies of the dense
matter equation of state \cite{J-MO2012,Owen2005,UCB2000}, nucleonic matter may
sustain deformations up to $\epsilon \simeq 10^{-6} - 10^{-7}$, whereas for
quark matter $\epsilon$ can reach $10^{-4} - 10^{-5}$. 

Observations of the majority of known pulsars show that their spin frequency
slowly decreases, $\dot{f} < 0$ (exceptions are pulsars in binary systems which
can be spun-up by the angular momentum transfer from the accretion disk). A
useful quantity related to the amount of kinetic (rotational) energy of the
star is the so-called spin-down limit. It is derived by assuming that the
gravitational-wave emission alone is responsible for the change in the
rotational energy, $\dot{E}_{\rm rot}$. For $E_{\rm rot} = 2\pi^2If^2$,
$\dot{E}_{\rm rot}\propto If\dot{f\ }$ is equated with the GW emission,
$\dot{E}_{\rm GW}\propto \epsilon^2 I^2 f^6$ to obtain the spin-down limit
amplitude
\begin{eqnarray} 
h_{\rm sd} = \frac{1}{d}\sqrt{\frac{5GI}{2c^3}\frac{|\dot{f\ }|}{f}} = 
8\times 10^{-24} \sqrt{\left(\frac{I}{10^{45}\ {\rm g\ cm^2}}\right)
\left(\frac{|\dot{f\ }|}{10^{-10}\ {\rm Hz/s}}\right)
\left(\frac{100\ {\rm Hz}}{f}\right)}
\left(\frac{100\ {\rm pc}}{d}\right).
\label{eq:h0sd} 
\end{eqnarray}
Comparison with Eq.~\ref{eq:h0} results in the limiting deformation $\epsilon_{\rm sd}$:
\begin{equation}
\epsilon_{\rm sd} = 2\times 10^{-5}
\sqrt{\left(\frac{10^{45}\ {\rm g\ cm^2}}{I}\right)
\left(\frac{100\ {\rm Hz}}{f}\right)^5
\left(\frac{|\dot{f\ }|}{10^{-10}\ {\rm Hz/s}}\right)}  
= 0.2\left(\frac{h_{\rm sd}}{10^{-24}}\right)f^{-2}I_{45}^{-1}d_{kpc}.
\end{equation}

A star with $\epsilon_{\rm sd}$ would spin-down solely by gravitational-wave
radiation.  In reality, the ellipticity is smaller, so the results `beating the
spin-down limit' probe the physically interesting range of ellipticities and
set the upper limit for the ellipticity given object has. We will come back to
the spin-down limit in Sect.~\ref{sect:curr}. 

\begin{figure}[h]
\begin{minipage}{0.5\linewidth}
\centerline{\includegraphics[width=\linewidth]{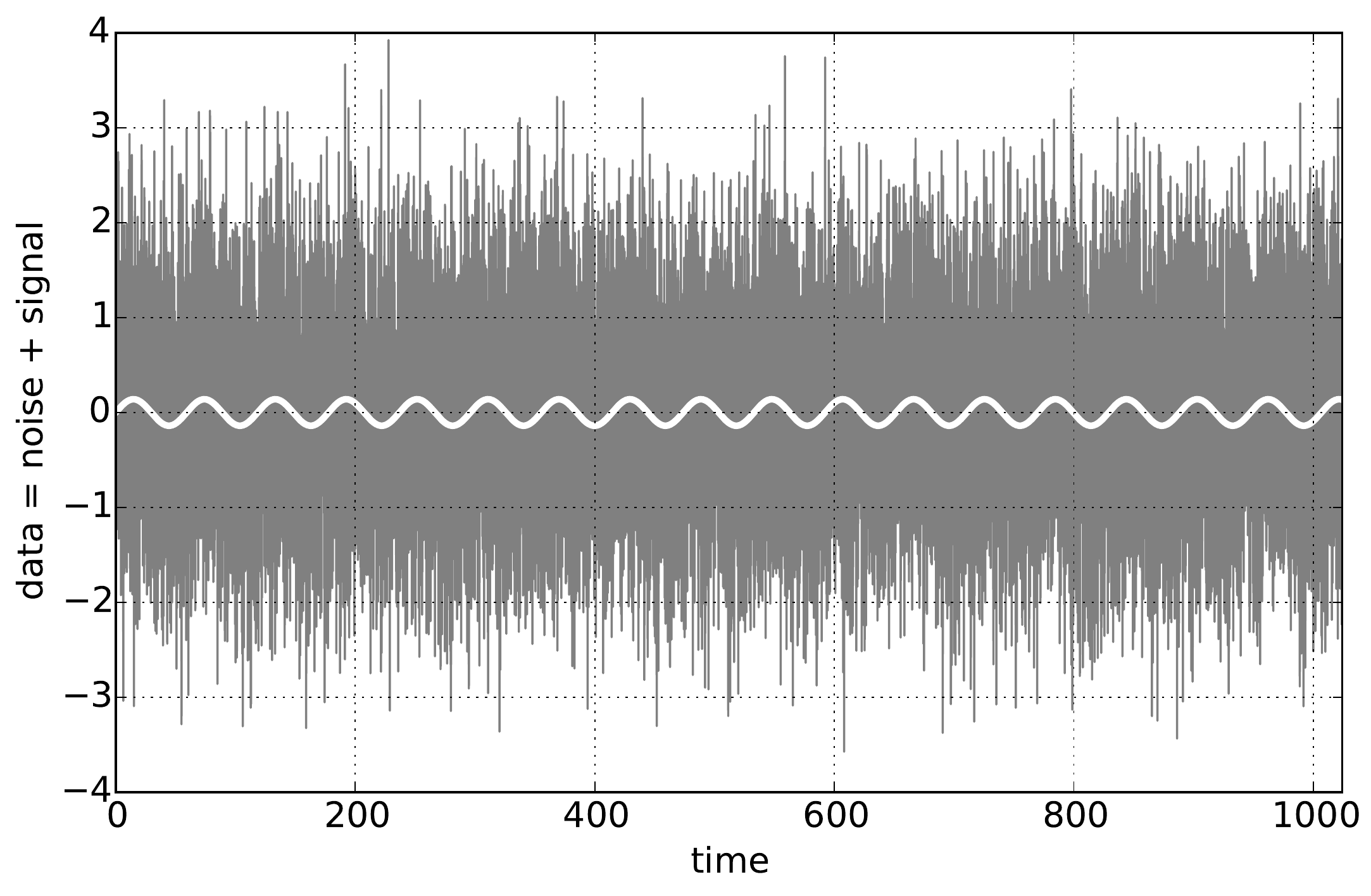}}
\end{minipage}
\hfill
\begin{minipage}{0.5\linewidth}
\centerline{\includegraphics[width=\linewidth]{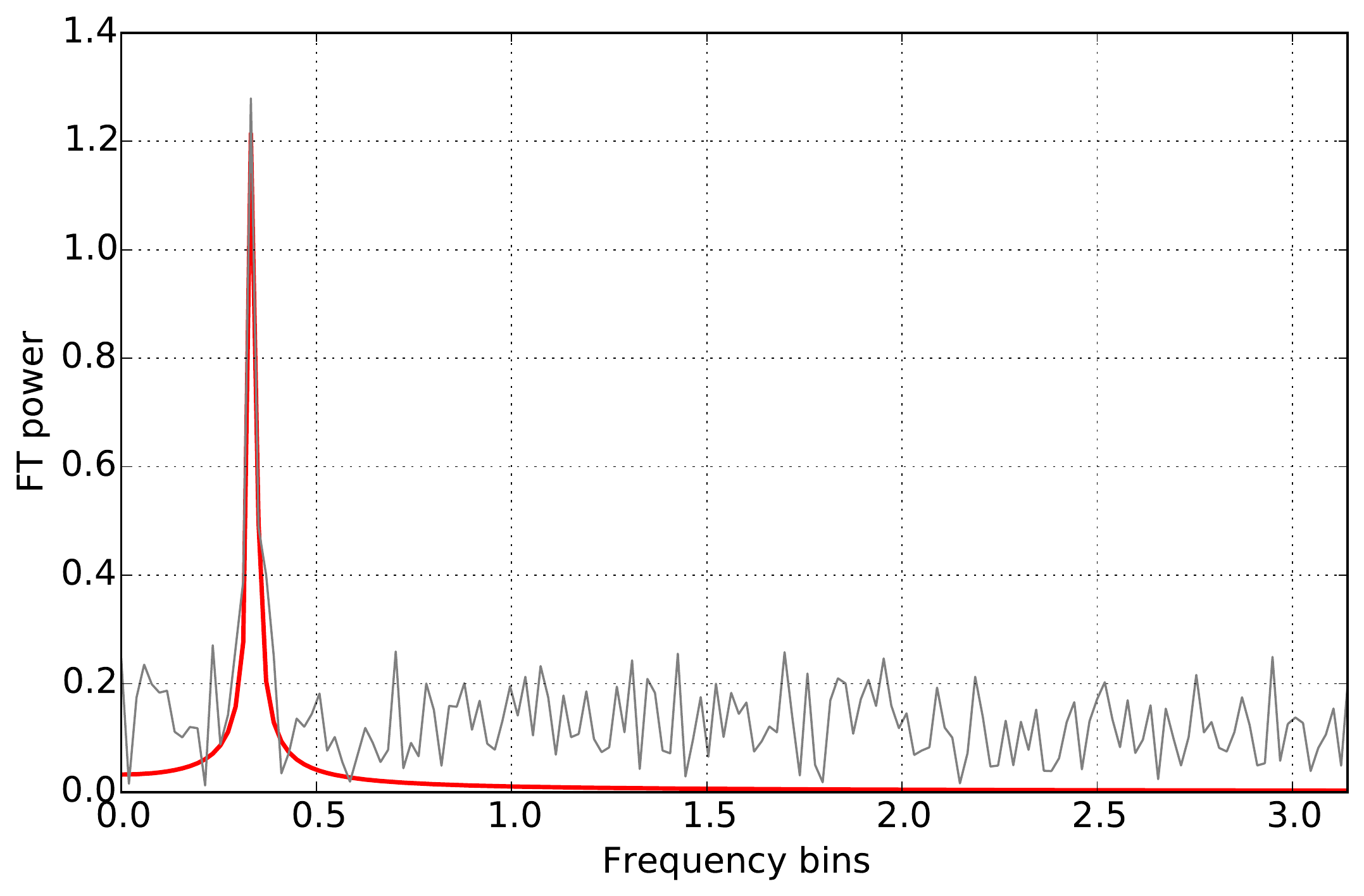}}
\end{minipage}
\caption{Periodic signal hidden in white Gaussian noise (left), and the 
Fourier transform of this time series (right).}
\label{fig:fft}
\end{figure}

\section{Data-analysis methods and computational challenges} \label{sect:meth} 

Searching for long-lived, weak gravitational wave signals is a particularly
cumbersome task, especially when nothing is known about its parameters.
Eq.~\ref{eq:h0} shows that their strain amplitude is much smaller than than
e.g., the characteristic strain amplitude of O1 detections, $h\simeq 10^{-21}$.
Fortunately, one may search for weak signals hidden deeply in the noise
provided a waveform of the signal is known. The idea of such a search would be
to compute the cross correlation between the data and a parametrised template
waveform in order to find parameters of the best match. Techniques of this sort
are called the matched filtering methods \cite{Wiener1949,Helstrom1968,Schutz1999}. 
Matched filtering provide an optimal detection statistic (it maximizes the signal-to-noise ratio) 
if noise is Gaussian. 

\subsection{Example: signal-to-noise ratio for a periodic signal} 

An instance of a strictly periodic signal ''buried'' in the stationary white
Gaussian noise is presented in Fig.~\ref{fig:fft}. We will estimate the
signal-to-noise ratio $\rho$ by approximating the output of the matched filter,
which in the case of a periodic signal is simply the Fourier transform. The
scalar product $(x\lvert y)$ is defined using the Fourier transform as 
\begin{equation} 
(x\lvert y) = 4\Re\int_{0}^{\infty}\frac{\tilde{x}(f)\tilde{y}^\ast(f)}{S_n(f)} \mathrm{e}^{2\pi ift} df, 
\end{equation} 
where $*$ denotes complex conjugation, $\Re$ is the real part of the
integral, and $\tilde{x}(f)$ and $\tilde{y}(f)$ are the Fourier transform of
the time-domain data series:  
\begin{equation}
  \tilde{x}(f) = \int_{-\infty}^\infty x(t) \mathrm{e}^{-2\pi ift} dt, 
\quad \text{with the inverse transform}\quad 
  x(t) = \int_{-\infty}^\infty \tilde{x}(f) \mathrm{e}^{2\pi ift} df. 
\end{equation} 
$S_n(f)$ is the one-sided power spectral density of the detector's noise. For a
stable detector we may assume the that $S_n(f)\approx S_0=const.$ over the data
span. From the Parseval theorem, 
\begin{equation}
  \left(x\lvert y\right) \approx \frac{2}{S_0}\int_0^T x(t) y(t) dt.
\end{equation}
For the additive noise process, the data $s(t)$ is defined as the sum of the
signal and the noise: $s(t) = h(t) + n(t).$ The matched filter output of the
data stream $s(t)$ with a filter template $h_{templ.}(t)$ (correlation of the
data containing a possible signal with its model) is 
\begin{equation} 
  4\Re\int_{0}^{\infty}\frac{\tilde{s}(f)\tilde{h}^\ast_{templ.}(f)}{S_n(f)} \mathrm{e}^{2\pi ift} df.
\end{equation}
The optimal signal-to-noise ratio is defined as $\rho := \sqrt{(h\lvert h)}$.
For a periodic signal $h(t) = h_0\cos(\phi(t) + \phi_0)$, we will assume that
the data span $T_0$ is much longer than the period of the wave, $P_0 = 1/f_0$,
and that the phase can be expanded in the series $\phi(t)=\Sigma_{k} a_k
t^{k+1}$. Also 
\begin{equation} 
  \frac{1}{T_0}\int_0^{T_0} \cos(n\phi(t))dt \approx \frac{1}{T_0}\int_0^{T_0} \sin(n\phi(t))dt \approx 0
\end{equation} 
for integers $n>0$. Integrating the $\rho^2$ for $h(t) = h_0\cos(\phi(t) + \phi_0)$ gives the 
estimate for the optimal signal-to-noise ratio for a periodic signal of the amplitude $h_0$ 
\begin{equation} 
  \rho = \left(\frac{2}{S_0}\int_0^{T_0} \left(h(t)\right)^2 dt\right)^{1/2} 
  \left( \frac{2}{S_0}\int_0^{T_0} h_0^2\cos^2(\phi(t) + \phi_0) dt\right)^{1/2} 
  \approx\ h_0 \left(\frac{T_0}{S_0}\right)^{1/2}. 
\end{equation} 
It is clear that even for a small $h_0$ one can reach a satisfactory signal-to-noise ratio 
with a sufficiently long stretch of data. 

In practice, on top of the secular spin-down modulation mentioned before (to describe
this feature, one says that the signal is {\em almost monochromatic}), the signal
is modulated by the movement of the detector with respect to the source. Since
the Advanced LIGO and Advanced Virgo detectors are placed on Earth, the
presence of other planets and Earth's rotation influences signal's amplitude
and phase. Demodulation to a frame connected with the Solar System Barycenter 
(the place in the Solar System that moves the least with respect to the source), 
precise ephemerides of the movement of planets are used. The fact that the 
detectors are moving with respect to the source isn't necessary a bad thing, 
though: detector movement distinguishes a real signal from local spectral 
artifacts, called the ''stationary lines''. 

\subsection{Example: the ${\mathcal{F}}$-statistic} 

A conceptually relatively simple method to develop a detection statistic using
the time-domain data $s(t)$ is the ${\mathcal{F}}$-statistic
\cite{JaranowskiKS1998}. For a triaxial rotating neutron star model
(Eq.~\ref{eq:triaxial}, the statistic is obtained by maximizing the likelihood
ratio function with respect to the four unknown parameters: $h_0$, $\phi_0$,
$\iota$, and $\psi$. This leaves, in case when only first derivative of $f$
is taken into account, a function of four parameters: $f$, $\dot{f}$, $\alpha$ 
and $ \delta$. These four parameters form a parameter space in which the
signal's best match will be searched for. Assuming that the observation time
$T_0$ is the integer multiple of sidereal days, and that the bandwidth is
narrow (so that the spectral density of the noise $S_0$ is constant), the
${\mathcal{F}}$-statistic is evaluated \cite{Astone2010p5} as 
\begin{equation}
  {\cal F} = \frac{2}{S_0 T_0}\left(\frac{|F_a|^2}{\langle a^2\rangle} +
  \frac{|F_b|^2}{\langle b^2\rangle}\right), 
\label{eq:fstat} 
\end{equation} 
with  
\begin{equation}
  F_a = \int_0^{T_0} s(t) a(t) \exp(-i \phi(t)) dt,\quad F_b = \dots,\quad 
  \langle a^2\rangle = \int_0^{T_0} a(t)^2 dt,\quad \langle b^2\rangle = \dots, 
\end{equation}
$F_a$ and $F_b$ being the generalizations of the Fourier transforms from the
previous section. Amplitude modulation functions $a(t)$ and $b(t)$ are related
to detector's antenna response ($F_+ = a(t)\cos 2\psi + b(t)\sin 2\psi$,
$F_\times = -a(t)\sin 2\psi + b(t)\cos 2\psi$) and depend on the sources' sky
position $\alpha$ and $\delta$, and the phase modulation function $\phi(t)$
depends also on the frequency $f$ and spin-down of the source,
$\dot{f}$. The signal-to-noise ratio $\rho$ is related to $\mathcal{F}$ as follows: 
$\rho = \sqrt{2({\cal F} - 2)}$.

\subsection{Taxonomy of search methods}

Continuous gravitational-wave searches can by divided according to the amount
of information one has about the sources. 

The {\em targeted searches} are most
often based on matched filtering (data of length $T_0$ correlated with signal
templates). Position, $f$ and $\dot{f}$, sometimes also the source's
orientation are known. In this case the expected strain amplitude scales like
$h_0 \propto \sqrt{S/T_0}$, where $S$ is the amplitude spectral density at the
expected gravitational-wave frequency. 

{\em Directed searches} cover the intermediate cases when only some parameters
are known, e.g., the position of the source. Astrophysically, they may be
relevant to supernov{\ae} remnants, the Galactic center, globular clusters,
accreting neutron stars in binary systems (e.g., the brightest X-ray source in the sky, 
Sco X-1). 

{\em All-sky searches} are the most demanding types of searches. Source
parameters and positions are not known, which makes the parameter space large
and the problem becomes very quickly computationally bound. In order to mediate
this, hierarchical approaches are being used. Instead of analyzing the whole
$T_0$ data span at once, the data is divided into $N$ data segments of length
$T_s$, which are analyzed coherently, and the resulting information is combined
incoherently. The expected strain amplitude scales like $h_0 \propto
\sqrt{S/T_s}/N^{1/4}$. The most sophisticated example of the hierarchical
approach is the volunteer-driven Einstein@Home project \footnote{{\tt
https://einsteinathome.org}}. 
  
\subsection{Example: computational cost for an all-sky search} 

In order to optimally cover the $(f, \dot{f}, \alpha, \delta)$ parameter space
of an all-sky search at all possible frequencies, a grid of parameters is
obtained as a solution to the covering problem with constraints (a constraint
being e.g., a condition that the optimal $(\dot{f}, \alpha, \delta)$ lattice
coincides with points in $f$ corresponding to the Fourier frequencies bins of 
the Fast Fourier Transform algorithm). Typically, the number of points in $\dot{f},
\alpha, \delta$ scales with some positive power of $T_0$, so, depending 
on the details of the implementation \cite{Wette2014}, the computational demand scales like 
\begin{equation}
\underbrace{T^2_0}_{\dot{f}}\times\underbrace{T^{[0-3]}_0}_{\alpha,\delta}\times \underbrace{T_0\log(T_0)}_{f\ \text{by FFT}} = T_0^{[3-6]}\log(T_0),  
\end{equation}
which is very prohibitive for large $T_0$. A coherent search of $T_0\simeq
1\,yr$ of data (comparable with ongoing and planned Advanced LIGO/Virgo runs)
in the broad sensitivity band of the detectors (10 - 2000 Hz) requires {\em
zettaFLOPS} ($10^{21}$ FLOPS) scale supercomputers. The solution is a
hierarchical scheme: divide the data $T_0$ into a number $N$ shorter length
$T_s$, $T_s\simeq$ days, data segments and perform a coherent search in each of
them (search in narrow frequency bands of bandwidth $B$, Nyquist sampling time
$\delta t = 1/2B$, number of data points $N_p = T_s /\delta t = 2T_s B$). 
\begin{figure}[h]
\begin{minipage}{\linewidth}
\centerline{\includegraphics[width=0.4\linewidth]{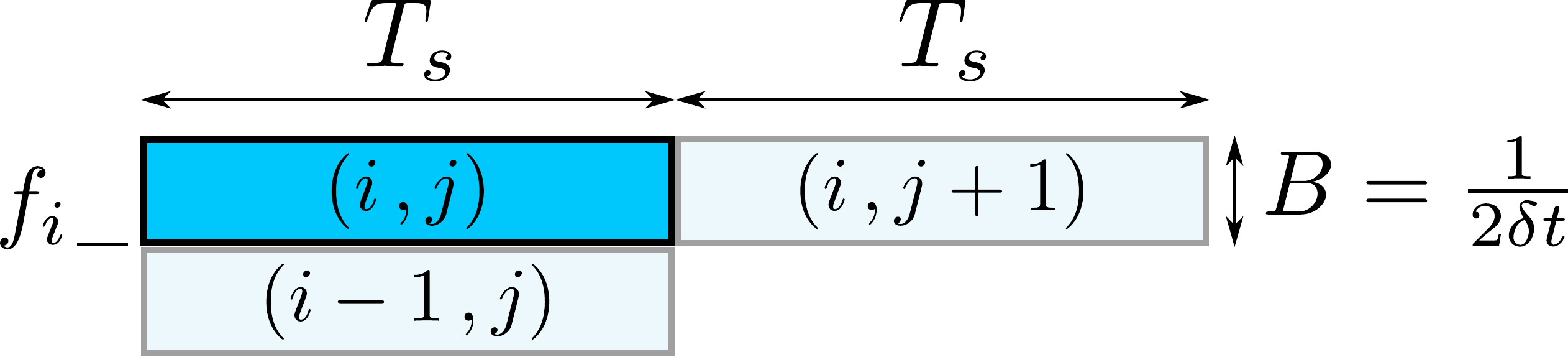}}
\end{minipage}
\caption{Division of the time-frequency data into narrow frequency bands of bandwidth $B$, 
reference frequency $f_i$, and $N$ time segments of length $T_s$ ($NT_s = T_0$). Each 
time-frequency segment is then analyzed separately, allowing for effective parallelization.} 
\label{fig:time_freq}
\end{figure}
This is feasible on a typical petaFLOP scale supercomputer (cluster), yet still
requires millions of CPU-hours. Second incoherent stage consists of searching
for coincidences between different $T_s$ segments. Surviving outliers present
in sufficiently many segments (candidate signals with relatively
well-determined parameters) are subjected to a final scrutinizing follow-up (a
''targeted search''). 

\section{Continuous gravitational waves in the Advanced LIGO O1 era} \label{sect:curr} 

The first Advanced LIGO observing run (O1) started on September 11, 2015 and
finished January 19, 2016. During that time the Hanford and Livingston
detectors collected 78 days and 66 days of science data, respectively. 

First published result pertains to a targeted search for gravitational waves
from 200 known pulsars \cite{Abbott2017a}. In this list, 11 high-value target
pulsars, for which the spin-down limit based on Eq.~\ref{eq:h0sd} could either
be improved or closely approached, were identified. These selected pulsars were
analyzed by three, largely independent methods: two time-domain-based methods,
Bayesian \cite{DupuisW2005} and $\mathcal{F}$/$\mathcal{G}$-statistic
\cite{JaranowskiK2010}, and the frequency-domain-based 5n-vector method
\cite{Astone2010,Astone2012}. The remaining 189 targets were analyzed using the
Bayesian method only. The analysis didn't find significant evidence for a
gravitational-wave signal from any of these pulsars, but the most constraining
upper limits to date on their gravitational-wave amplitudes and ellipticities
were obtained. For eight of the high-value target pulsars, new upper limits
give improved bounds over the indirect spin-down limit values. For the Crab
pulsar, the 95\% confidence upper limit for the gravitational-wave radiation
energy is $2\times 10^{-3}\ \dot{E}_{rot}$, and in case of the Vela pulsar, the
upper limit is $10^{-2}\ \dot{E}_{rot}$. The limits on ellipticity correspond
to the relative deformation (the ''mountain'') no greater than $\simeq 10$ cm
and $\simeq 50$ cm for the Crab and Vela pulsars, respectively. For
another 32, values within a factor of 10 of the spin-down limit were found: it
is likely that some of these will be reachable in future runs of Advanced LIGO
and Advanced Virgo. The smallest upper limit was obtained for PSR J1918-0642:
$h_0 = 1.6\times 10^{-26}$. These new results improve on previous limits of
Initial LIGO/Virgo by more than a factor of two (see the summary Fig. 1 of
\cite{Abbott2017a}, where the comparison with the sensitivity curve, initial
detector results and the spin-down limits is presented). 

Second publication (in the time of writing available as a preprint) is related
to a directed search for gravitational waves from a bright X-ray source Sco X-1
\cite{Abbott2017b}. Sco X-1 is the brightest Low Mass X-ray Binary (LMXB, a
binary system consisting of a neutron star or a black hole, and a normal star
with lower mass) in the Galaxy. The X-ray radiation is produced during
accretion; neutron stars in these systems are potential sources of continuous
gravitational waves because accretion provides a natural method of building a
deformation on the star and powering the gravitational-wave emission. There are
however challenges in searching for gravitational waves from this particular
source. First, the spin frequency of the neutron star is unknown - the search
has to cover a broad range of frequencies, which means it requires much more
computing power than a directed search in case of a known pulsar. Most likely 
the spin frequency is not constant, and not even behaving
strictly monotonously, but ''wandering'' i.e., it changes because of the
fluctuations in the amount of accreted matter; frequency is also modulated by
the orbital motion of the binary system (the signal power is distributed into
sidebands i.e., frequencies higher or lower than the gravitational-wave signal
frequency). In order to perform an efficient search in these circumstances, a
hidden Markov model (a statistical model in which it is assumed that the system
is a Markov process with unobserved states, see \cite{Baum1972} and references
therein and tutorial \cite{Rabiner1989}) was implemented. The search covered a
band of frequencies from 60 Hz to 650 Hz. No detection was claimed from this
search, but very sensitive upper limits on the gravitational-wave strain were
placed (to quote one example: 95\% upper limits $h_0 = 3\times 10^{-25}$ at 100 Hz, 
assuming circular polarization).  

We also report on an all-sky search for periodic gravitational waves
using the Advanced LIGO's O1 run \cite{O1CWAllSky}, in the frequency band of 20-475\,Hz and a
frequency time derivative range of $[-1.0, +0.1]\times 10^{-8}$\,Hz/s. Several
different data-analysis pipelines took part in this study: the {\it PowerFlux}
(see \cite{S6PowerFlux} and references therein), the {\it FrequencyHough}
\cite{VSRFH}, the {\it Skyhough} \cite{S5Hough} and the {\it Time-Domain
$\mathcal{F}$-stat} \cite{VSR1TDFstat}. The pipelines employ a variety of
algorithmic and parameter choices e.g., they primarily use either the frequency
or time domain data, adopt different coherence times used in first-stage data
processing (from 1800 s to 6 days), and treat differently the narrow spectral
artifacts (''lines''). 
\begin{figure}[h]
\begin{minipage}{\linewidth}
\centerline{\includegraphics[clip,trim=0 {0.02\textheight} 0 {0.07\textheight},width=0.8\linewidth]{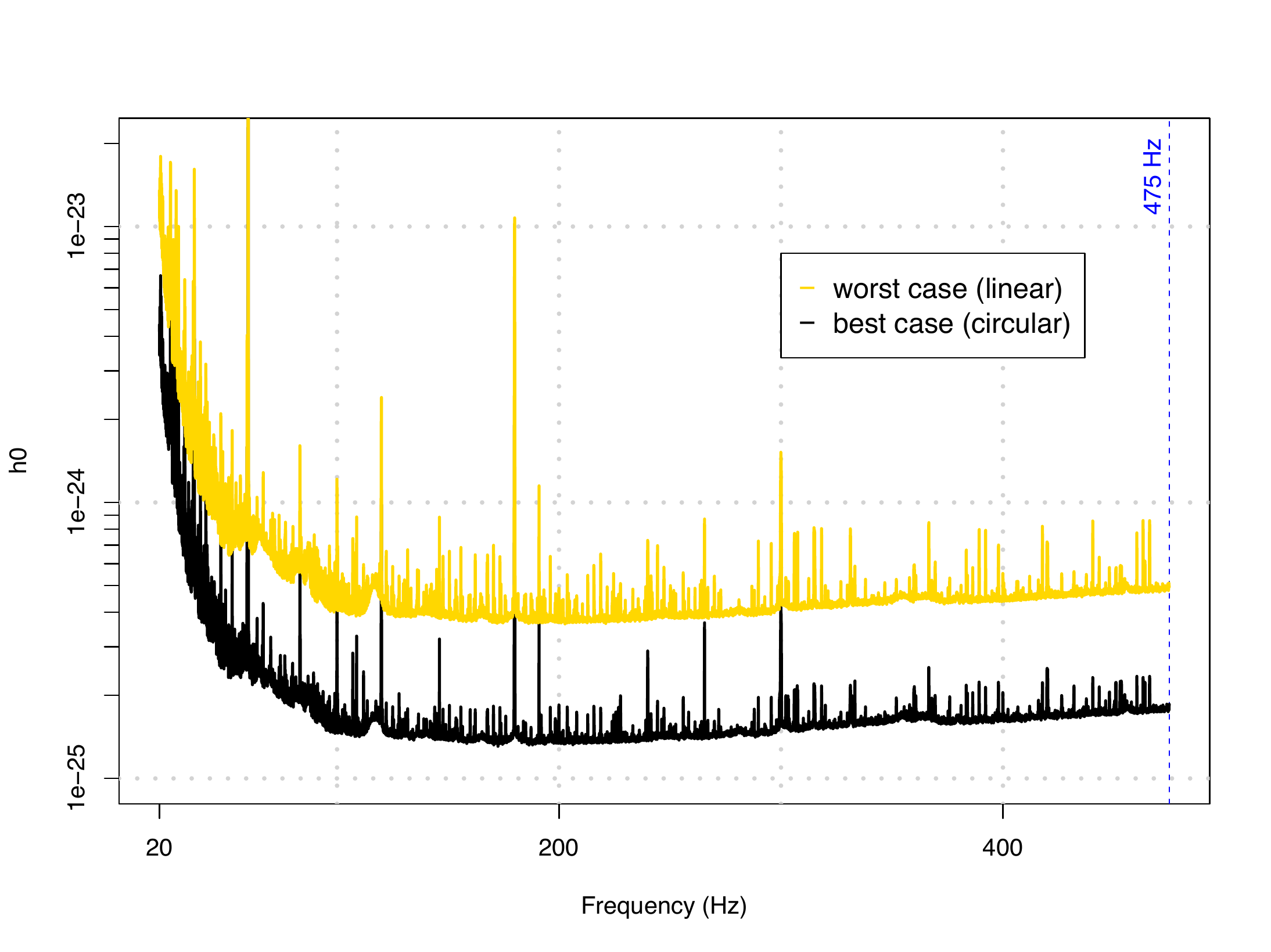}}
\end{minipage}
\caption{O1 all-sky upper limits for continuous gravitational waves from the {\em PowerFlux} pipeline. The upper (yellow) curve shows worst-case (linearly polarized) $95$\%~CL upper limits. The lower (grey) curve shows upper limits assuming a circularly polarized source.} 
\label{fig:o1allsky}
\end{figure}
Outliers that survive all stages of any of the four pipelines are examined
manually for contamination from known, or possibly new, instrumental artifacts. 
Survivors of this procedure are subjected to additional systematic
follow-up used for Einstein@Home searches \cite{ehfu}. Upper
limit results are presented in Fig.~\ref{fig:o1allsky}. The lowest upper
limits on worst-case (linearly polarized) strain amplitude $h_0$ are  $\simeq 
4\times 10^{-25}$ near 170\,Hz. For a circular polarization (most favorable
orientation of the source), the smallest upper limits obtained are $\simeq 
1.5\times 10^{-25}$.

In addition, several studies from the Initial Detector Era (LIGO S6 and Virgo VSR2, VSR4
runs) were recently published. They include an all-sky Mock Data Challenge
based on the LIGO S6 data \cite{AllSkyMDC}, a directed search towards nine
supernova remnants \cite{S6SNRPaper}, the Orion spur \cite{orionspur}, the
globular cluster NGC 6544 \cite{S6GlobularCluster}, as well as the deepest
all-sky survey for continuous waves from the Initial Detector Era, the S6
Einstein@Home search in the $[50, 505]$ Hz range \cite{EAHS6}. The reason we
see merit in publishing the `old' results is to gather experience, test
algorithmic improvements and develop new methods using the well-understood
data. These new tools are now being used in the O1 (and soon, the O2)
searches. 

\section{Plans for the future} \label{sect:plans} 

The algorithmic and implementation-related improvements acquired during the
Dark Ages (2011--2015) will be used to expand the `standard' targeted searches
for high-value targets (Sco X-1, Cas A, Vela Jr and G347 supernov{\ae}
remnants, Crab and Vela pulsars) and to speed up the massive all-sky searches
with relatively simple source models (aligned triaxial ellipsoid,
Eq.~\ref{eq:triaxial}). Additionally, we plan to search for signals with more
complicated, realistic morphology. The models include inclined rotating neutron
stars, which emit gravitational waves at multiple frequencies at once (e.g., at
$f$ and $2f$), transient continuous gravitational waves emitters (phenomena
that may last for weeks to months, and are caused by neutron-star instabilities
e.g., the r-modes), and search for non-tensorial gravitational waves. In order
to capture the richness of physical processes will are also improving the
loosely-coherent methods taking into account the neutron-star frequency
wandering, glitches, and a possible mismatch between the gravitational-wave spin
frequency parameters and the parameters inferred from the electromagnetic
observations. 

\section*{Acknowledgments}
This work was partially supported by the Polish NCN grant no. UMO-2014/14/M/ST9/00707 
and ASPERA/NCN grant 2013/01/ASPERA/ST9/00001.


\section*{References}

{\small 
\begin{thebibliography}{99}

\bibitem{GW150914} Abbott, B.~P., Abbott, R., Abbott, T.~D., et al.\ 2016, \prl, 116, 061102 

\bibitem{GW151226} Abbott, B.~P., Abbott, R., Abbott, T.~D., et al.\ 2016, \prl, 116, 241103 

\bibitem{ALIGO2015} Aasi, J., Abbott, B.~P., et al.\ 2015, \cqg, 32, 074001

\bibitem{AdV2015} Acernese, F., Agathos, M., Agatsuma, K., et al.\ 2015, \cqg, 32, 024001 

\bibitem{AnderssonK2001} Andersson, N., \& Kokkotas, K.~D.\ 2001, IJMP D, 10, 381 

\bibitem{Lasky2015} Lasky, P.~D.\ 2015, PASA, 32, e034 

\bibitem{Einstein1918} Einstein, A.\ 1918 ''Sitzungsberichte der K{\"o}niglich Preussischen Akademie der Wissenschaften zu Berlin'', 154

\bibitem{J-MO2012} Johnson-McDaniel, N.~K., \& Owen, B.~J.\ 2012, \prd, 86, 063006 

\bibitem{Owen2005} Owen, B.~J.\ 2005, \prl, 95, 211101 

\bibitem{UCB2000} Ushomirsky, G., Cutler, C., \& Bildsten, L.\ 2000, \mnras, 319, 902

\bibitem{Wiener1949} Wiener, N.\ 1949, ''Extrapolation, Interpolation, and Smoothing of Stationary Time Series'', New York: Wiley 

\bibitem{Helstrom1968} Helstrom, C. W.\ 1968, ''Statistical Theory of Signal Detection'', Pergamon Press, London

\bibitem{Schutz1999} Schutz, B.~F.\ 1999, \cqg, 16, A131 

\bibitem{JaranowskiKS1998} Jaranowski, P., Kr{\'o}lak, A., \& Schutz, B.~F.\ 1998, \prd, 58, 063001 

\bibitem{Astone2010p5} Astone, P., et al.\ 2010, \prd, 82, 022005 

\bibitem{Wette2014} Wette, K.\ 2014, \prd, 90, 122010 

\bibitem{Abbott2017a} Abbott, B.~P., Abbott, R., Abbott, T.~D., et al.\ 2017, \apj, 839, 12

\bibitem{DupuisW2005} Dupuis, R. J., Woan, G. 2005, \prd, 72, 102002

\bibitem{JaranowskiK2010} Jaranowski, P., Kr{\'o}lak, A. 2010, \cqg, 27, 194015

\bibitem{Astone2010} Astone, P., D’Antonio, S., Frasca, S., Palomba, C. 2010, \cqg, 27, 194016

\bibitem{Astone2012} Astone, P., et al., 2012, JPCS, 363, 012038

\bibitem{Abbott2017b} Abbott, B.~P., et al.,\ 2017, \prd, 95, 122003  

\bibitem{Baum1972} Baum, L.E.,\ 1972, Inequalities, 3, 1

\bibitem{Rabiner1989} Rabiner, L.,\ 1989, Proceedings of the IEEE, 77, 257

\bibitem{O1CWAllSky} Abbott, B.~P., et al.\ 2017, \prd, 96, 062002

\bibitem{S6PowerFlux} Aasi, J. et al., 2016,\ \prd, 94, 042002

\bibitem{VSRFH} Aasi, J. et al., 2016,\ \prd, 93, 042007

\bibitem{S5Hough} Aasi, J. et al., 2014,\ \cqg, 31, 085014

\bibitem{VSR1TDFstat}  Aasi, J. et al., 2014,\ \cqg, 31, 165014

\bibitem{ehfu} Papa, M. A., et al., 2016,\ \prd,94, 122006

\bibitem{AllSkyMDC} Walsh, S. et al., 2016,\ \prd, 94, 124010

\bibitem{S6SNRPaper} Aasi, J. et al., 2015,\ \apj, 813, 1

\bibitem{orionspur} Aasi, J. et al., 2016,\ \prd, 93, 042006 

\bibitem{S6GlobularCluster} Abbott, B.~P., et al.,\ 2016, \prd, 95, 082005 

\bibitem{EAHS6} Abbott, B.~P., et al.\ 2016, \prd, 94, 102002

\end{thebibliography}

}
\end{document}